\documentclass[aps,nofootinbib,twocolumn]{revtex4}

\usepackage{amssymb,amstext,amsmath,amsthm}
\usepackage[dvips]{graphicx}
\usepackage{latexsym}
\usepackage{psfrag}
\usepackage{amsfonts}
\usepackage{bbm}

\newcommand{\Ref}[1]{(\ref{#1})}

\def\nn{\nonumber}

\newcommand{\eqa}{\begin{eqnarray}}
\newcommand{\neqa}{\end{eqnarray}}
\newcommand{\equ}{\begin{equation}}
\newcommand{\nequ}{\end{equation}}
\newcommand{\no}{\nonumber\\}

\def\w{\wedge}

\newcommand{\p}{\partial}

\def\th{\theta}
\def\al{\alpha}

\def\f{\frac}
\def\tl{\tilde}

\usepackage{bbm}

\let\eps=\epsilon

\let\si=\sigma

\newcommand{\rs}{{\rm s}}
\newcommand{\rc}{{\rm c}}

\def\tl{\tilde}

\begin{document}

\title{\Large\bf Area-angle variables for general relativity}

\author{Bianca Dittrich and Simone Speziale}
\affiliation{Perimeter Institute, 31 Caroline St. N, Waterloo, ON N2L 2Y5, Canada.}
\date{\small\today}

\begin{abstract}
We introduce a modified Regge calculus for general relativity on a triangulated four dimensional 
Riemannian manifold where the fundamental variables are areas and a certain class of angles.
These variables satisfy constraints which are local in the triangulation.
We expect the formulation to have applications to classical 
discrete gravity and non-perturbative approaches to quantum gravity.
\end{abstract}

\maketitle

\section{Introduction}
Discrete approaches have proved useful in many areas of physics. 
Regge calculus \cite{Regge} is a discrete formulation of general relativity (GR) where
spacetime is approximated by a triangulated manifold, and the fundamental variables used to describe the metric
are the lengths of the edges of the triangulation. This approach
has been applied with some success to classical gravity \cite{RuthRegge, Gentle}, and used as a starting point for a
lattice quantization of GR \cite{RuthRegge, Rocek, Immirzi, Hamber}. 
As other non-perturbative approaches to quantum gravity, 
quantum Regge calculus suffers from the problem of defining a unique gauge-invariant measure in the path integral.
The background-independent spinfoam approach \cite{carlo} suggests an original route based on the
well-defined quantum measure of BF theory. The latter is a topological theory where area variables 
appear naturally, and whose action can be reduced
to GR by means of so-called simplicity constraints, as discovered by Plebanski \cite{Plebanski}. 
This and other motivations have led Rovelli \cite{carloarea} to suggest that 4d quantum gravity should be
related to a modification of Regge calculus where the fundamental variables are the areas 
of triangles rather than the edge lengths.
Some effort was put in this line of research by Makela and Williams among others \cite{Makela,Barrett,Wainwright,RuthRegge},
but the problem has been open for more than ten years. 
The main difficulty lies in the fact that a generic triangulation has many more triangles than edges,
thus area variables should be constrained. An
explicit expression of these constraints is obscured by their non-local nature in the triangulation.

In this paper, we introduce a description of discrete gravity that overcomes this difficulty.
The key idea is to enlarge the set of variables from areas only, to areas and angles. 
In this way the constraints become local, are easy to write explicitly, and further they are related to the 
simplicity constraints of Plebanski's formulation of GR.\footnote{Similar ideas have been
investigated by Reisenberger \cite{Mike} and Rovelli \cite{CarloUn}.}

We approximate the spacetime manifold by a simplicial triangulation, where each 4-simplex\footnote{The 
4-simplex, also known as pentachoron in the mathematical literature,
is the convex hull of five points. A 4-simplex contains five tetrahedra, ten triangles and ten edges.}
is flat and the curvature is described by deficit angles associated to the triangles. 
Regge calculus uses the fact that on each 4-simplex $\sigma$ the ten components of the (constant) metric tensor $g_{\mu\nu}(\sigma)$ can be straighforwardly expressed in terms of the ten edge lengths $\ell_e$. 
A further advantage of using the edge lengths as variables is that they
endow each tetrahedron with six quantities which are sufficient
to completely characterize the tetrahedron's geometry. Therefore the gluing of 4-simplices, obtained
by identifying a shared tetrahedron, is trivial and causes no complications.

On a single 4-simplex, there are also ten triangles, suggesting that areas can be
equivalently taken as the metric variables. There are two difficulties with this idea.
First, it is less straighforward to express $g_{\mu\nu}(\sigma)$ in terms of areas.
For instance the change of variables from edge lengths to areas on a 4-simplex
is singular for orthogonal configurations \cite{Barrett}, that is where right angles among the edges
are present. 
Even the equal area configuration has a two-fold ambiguity where the same set of areas corresponds to
two different sets of edge lengths.
This is a more significant difficulty than it might seem at a first look, as such configurations are 
relevant in the case of a regular lattice, the simplest flat solution to Regge calculus.
The second issue is even more serious. 
Ten areas might be enough to describe the 4-geometry of the simplex, but how about
its boundary 3-geometry? Taken any of the five tetrahedra in the 4-simplex, its geometry is 
not uniquely defined by the areas of its 4 triangles
(two more quantities are needed, corresponding for instance to (non-opposite) dihedral angles).
So we need the geometry of the full 4-simplex to determine the individual geometry of any of 
its boundary tetrahedra. As a consequence, two adjacent 4-simplices in a triangulation will typically induce 
different geometries on the common tetrahedron, leading to discontinuities in the metric \cite{Wainwright},
or to non-local constraints involving the two 4-simplices \cite{Makela}.

A solution to the problem can be achieved adding to the areas
enough variables so that the geometry of each of the five tetrahedra can be independently and
completely determined. A natural choice is to add the tetrahedral dihedral angles.
Of course, this pleonastic set of variables needs to be constrained in order
to succesfully reproduce the dynamics of GR. We now turn to the study of these
constraints.

\section{De natura pentachori}
Let us study how to characterize the geometry of a 4-simplex and its five boundary tetrahedra, using 
areas and 3d dihedral angles.
We use a notation which might seem counterintuitive at first, but that pays off well in terms of
efficiency and extends to any dimension. 
We denote by $V$ the 4-volume of the simplex $\si$ (or the $n$-volume in general), $V(i)$ the 3-volume
of the tetrahedron $\si(i)$ obtained by removing the vertex $i$ from the 4-simplex, $V(ij)$ the area
of the triangle $\si(ij)$ obtained by removing the vertices $i$ and $j$, and so on.
For the dihedral angles, we use the following notation: $\th_{ij}$ is the 4d dihedral angle
between the tetrahedra $\si(i)$ and $\si(j)$, hinged at the triangle $\si(ij)$;
$\phi_{ij,k}$ is the 3d dihedral angle between the two triangles $\si(ik)$ and $\si(jk)$, 
hinged at $\si(ijk)$ within the tetrahedron $\si(k)$; finally, 
$\al_{ij,kl}$ is the 2d dihedral angle between the edges $\si(ijk)$
and $\si(ijl)$ belonging to the triangle $\si(kl)$.
All dihedral angles are \emph{internal}, thus for instance an equilateral
$4$-simplex has $\cos\th=1/4$.

\begin{figure}[h]
\includegraphics[width=2cm]{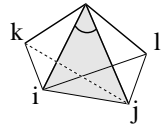}
\caption{\label{fig1} The geometric meaning of equation \Ref{aa}: the 2d angle $\al_{ij,kl}$ belonging to the shaded
triangle can be expressed in terms of 3d angles associated the thick edges of the tetrahedron $k$, or
equivalenty of the tetrahedron $l$.}
\end{figure}

These various types of dihedral angles satisfy a number of relations in a closed 4-simplex, 
which we present together with their proofs in the Appendix. 
An important role in our construction is played by the following expression of the
2d $\al$'s in terms of the 3d $\phi$'s,
\eqa\label{a}
\cos\al_{ij,kl}=\frac{\cos\phi_{ij,k}+\cos\phi_{il,k} \cos\phi_{jl,k}  }{\sin\phi_{il,k} \sin\phi_{jl,k}}.
\neqa
In this formula the 2d angle, belonging to the triangle $kl$, is described in terms of three 3d angles
all belonging to the \emph{same} tetrahedron $k$. In a closed 4-simplex, a triangle is shared by two tetrahedra,
thus there are two possible choices. Consistency of the two choices, i.e. $\al_{ij,kl}=\al_{ij,lk}$
(see Fig.\ref{fig1}), gives
\eqa\label{aa}
{\cal C}_{kl,ij}(\phi) &\equiv& 
\f{\cos\phi_{ij,k}+\cos\phi_{il,k} \cos\phi_{jl,k}  }{\sin\phi_{il,k} \sin\phi_{jl,k}} \no &-&
\f{\cos\phi_{ij,l}+\cos\phi_{ik,l} \cos\phi_{jk,l}  }{\sin\phi_{ik,l} \sin\phi_{jk,l}} = 0.
\neqa

Thus a consistent gluing of the tetrahedra in a 4-simplex gives relations among the $\phi$'s.
These are three relations per triangle, hence $30$ in total, of which only $20$ are independent.
To see this, we linearized the equations (\ref{aa}) around generic non-degenerate configurations, including the 
potentially harmful orthogonal one, and used an algebraic manipulator to study the rank. 
The good behaviour of the orthogonal configuration can 
also be anticipated by the absence of cosines in the denominator of \Ref{a}. 
Of course, our construction would fail for degenerate
configurations where one or more angles equal $0$ or $\pi$.

These relations are important to characterize the geometry of a 4-simplex.
Consider a generic 4-simplex. Its ten 4d angles $\th_{ij}$ define the Gram matrix
$G_{ij}(\th) \equiv \cos\th_{ij}$ (with the convention $\cos\th_{ii} \equiv -1$).
If the simplex is closed and flat, these ten angles can not be all independent, but have to
satisfy the condition of vanishing of the Gram determinant, $\det G=0$ 
(e.g. \cite{Barrett1} and \cite{Freidel}).\footnote{This
condition is the origin of the well-known Schl\"afli identity.}
The nine independent quantities parametrize the space of shapes of the 4-simplex
(a scale factor being the tenth and last metric variable).

We then expect that to characterize the geometry in terms of the thirty 3d angles $\phi_{ij,k}$, 
there must exist 21 relations among them. These can be found as follows.
First, we consider the Gram matrices $G^k{}_{ij}(\phi)$ associated to the five tetrahedra;
imposing the vanishing of their determinant guarantees that the tetrahedra are closed.
These are five independent conditions.
Next, we use the 2d angle consistency relations \Ref{aa} to ensure 
a consistent gluing of the tetrahedra into a 4-simplex.
The complete set 
\equ\label{C1}
\det \,G^k(\phi) = 0, \quad
{\cal C}_{kl,ij}(\phi) = 0 
\nequ
can be shown, again by linearization, to have rank 21.
Notice that the first constraint is local on each tetrahedron, unlike the second that involves two adjacent tetrahedra.
Hence we found a necessary and sufficient set of relations among 
the $\phi$ angles to be the 3d dihedral angles of a 4-simplex. 
Other sets are possible (see Appendix B); 
the advantage of this one is the transparency of its geometric meaning.

\section{Area-angle Regge calculus}

With the understanding of the geometry of a 4-simplex gained above, we now come to the main point of this paper:
describing the dynamics of general relativity on a discrete manifold, using areas and 3d angles as variables.
For simplicity, we consider here the case of a Riemannian manifold with no boundaries. The extension
to Lorentzian signature and to boundary terms will be discussed elsewhere.

Using the standard notation ($t$ a triangle, $e=tt'$ an edge), the variables 
on the full triangulation are $A_t$ and $\phi_e^\tau$.
The gluing conditions \Ref{aa} refer to each pair of edges in a triangle shared by two tetrahedra.
In a triangulation there will be in general many tetrahedra around the same triangle, and \Ref{aa} has to hold
for any choice of two.
However transitivity ensures that it is enough to impose \Ref{aa} to the pairs of tetrahedra belonging to
the same 4-simplex. We can then write these constraints as
\equ\label{Cee}
{\cal C}^\si_{ee'}(\phi_e^\tau)=0, 
\nequ
where ${\cal C}^\si_{ee'}$ is
given by \Ref{aa} for $e=\si(kli)$ and $e'=\si(klj)$ sharing a vertex in a 4-simplex $\si$, and zero otherwise.

On a single 4-simplex we have ten areas and thirty 3d angles, thus
we need thirty independent constraints to reduce the total number of variables to ten. 
The situation parallels the analysis we performed in the previous section.
We can still take the triangle gluing conditions \Ref{Cee} involving only $\phi$ angles, and
include the areas in the closure conditions for the five tetrahedra. 
Denoting $n_t$ the normal to a triangle, we have by definition $|n_t|^2 = A_t^2$ and
$n_t \cdot n_{t'} = - A_t A_{t'} \cos\phi_{tt'}^\tau$. The closure condition on a tetrahedron $\tau$ reads
$N_\tau \equiv \sum_{t\in\tau} n_t = 0$. 
By sequentially taking the scalar product of $N_\tau$ with the four $n_t$ 
we obtain four constraints,
\equ\label{closure}
{\cal N}_t^\tau(A,\phi) = A_t - \sum_{t'\neq t} A_{t'}\, \cos\phi_{tt'}^\tau = 0.
\nequ
Considering the five tetrahedra on the whole 4-simplex \Ref{closure} gives twenty constraints, to 
be added to the thirty constraints \Ref{Cee}. Again we studied the number of independent constraints 
by linearization, and found that the resulting system has rank 30 for a generic configuration and also for the orthogonal one. Consequently only ten of the 40 variables used are truly independent. This is consistent
with the kinematical degrees of freedom of discrete general relativity.
As shown explicitly in the Appendix, the forty variables $(A_t, \phi_e^\tau)$ satisfying these thirty independent relations determine completely the geometry of the 4-simplex \emph{and} of its five tetrahedra,
thus each tetrahedron has a well-defined geometry, and gluing 4-simplices causes no problems.
In other words, satisfying the constraints \Ref{Cee} and \Ref{closure} allows us to reconstruct uniquely
a set of edge lengths from the variables $(A_t, \phi_e^\tau)$.

We then consider the following action for general relativity,
\eqa\label{action}
&& S[A_{t},\phi_e^\tau,\lambda_t^\tau, \mu^\si_{ee'}] = \sum_{t} A_{t} \, \eps_{t}(\phi) + \\\nn
&& \qquad + \sum_\tau \sum_{t\in\tau} \lambda_t^\tau \, {\cal N}_t^\tau(A, \phi) + 
\sum_\si \sum_{ee'\in \si} \mu^\si_{ee'} \, {\cal C}^\si_{ee'}(\phi).
\neqa
The first term is just the Regge action with independent area-angle variables,\footnote{The 
deficit angles $\eps_t$ are given by the sum over 4d angles on the 4-simplices
sharing the triangle $t$, $\eps_t = 2\pi -\sum_{\si} \th_t^\si$. We describe in the Appendix how
to express them in terms of the $\phi$'s, or in terms of edge lengths as it is done in Regge calculus.} 
and the other terms are the constraints \Ref{closure} and \Ref{Cee}
imposed by the Lagrange multipliers $\lambda_t^\tau$ and $\mu_{ee'}^\si$.
As discussed above, they effectively reduce the set of variables $(A_t, \phi_e^\tau)$ to the edge
lenghts $\ell_e$, therefore \Ref{action} is equivalent to the conventional Regge action,
$S_{\rm R}[\ell_e] = \sum_t A_t(\ell_e) \, \eps_t(\ell_e)$.

Notice that our approach should not be seen as a first order formulation of Regge calculus
(see for instance \cite{Caselle, Barrett1}), 
because we are adding 3d dihedral angles (the $\phi$'s), not 4d dihedral angles 
(the $\th$'s): only the latter encode the extrinsic curvature of a 3d slice
and are thus conjugate to the areas.

The reader might wonder at this point whether \Ref{action} is a discretization of a continuum action for GR,
like Regge's is a discretization of the Einstein-Hilbert action $\int \sqrt{g} \, R$.
We argue that this is the case, the continuum avatar of \Ref{action} being Plebanski's action \cite{Plebanski}.
The latter is a modified BF action, which schematically reads $S = \int B \w F + \mu \, {\cal C}(B)$
(we refer the reader to the literature for more details \cite{Pleb}). 
The term ${\cal C}(B)$ is a set of constraints reducing topological BF theory to GR.
We are naturally led towards the interpretation of the first two terms of \Ref{action}
as a discretization of BF theory, with the closure constraint \Ref{closure} implementing
the Gauss constraint of the continuum BF action,
and the third term as the simplicity constraints.
Recall that Plebanski's constraints state that the bi-normal
$B$ to any triangle $\si(ij)$ must be simple, i.e. it must be the wedge product
of (any) two edge vectors. In our notation, $B_{ij} = \pm e_{ijk} \w e_{ijl}$ with $k$ and $l$ 
different from $i$ and $j$. Then if the closure and simplicity constraints are satisfied, they imply
\eqa\label{simpl1}
B_{ij}\cdot B_{ij} &=& e_{ijk}{}^2 \, e_{ijl}{}^2 - (e_{ijk}\cdot e_{ijl})^2 = \no &=&
V_{ijk}{}^2 V_{ijl}{}^2 \sin\al_{ijkl}^2 = \no &=& V_{ij}{}^2, \\\label{simpl2}
B_{ij}\cdot B_{ik} &=& 
e_{ijk}{}^2 \, (e_{ijl} \cdot e_{ikl}) - (e_{ijk}\cdot e_{ijl}) \, (e_{ijk} \cdot e_{ikl})  \no && \hspace{-1.8cm} =
V_{ijk}{}^2 \, V_{ijl} \, V_{ikl} \, \big( \cos\al_{il,jk}-\cos\al_{ij,kl} \cos\al_{ik,jl} \big) = \no &=& 
V_{ij}\, V_{ik} \, \cos\phi_{jk,i},
\neqa
thus using \Ref{simpl1} in \Ref{simpl2} gives
\equ\label{ainv}
\cos\phi_{jk,i} = \f{\cos\al_{il,jk}-\cos\al_{ij,kl} \cos\al_{ik,jl} }{\sin\al_{ij,kl} \sin\al_{ik,jl}}.
\nequ
This relation can be inverted to give \Ref{a} with \Ref{aa} holding (see the Appendix). 
Conversely if \Ref{aa} and \Ref{closure} are satisfied, we can proceed backwards and define
bi-normals satisfying the simplicity constraints.\footnote{Notice that 
the simplicity constraints have typically solutions in two sectors. Here we are imposing directly
the solution in the geometric sector, so we do not comment about this ambiguity, which however plays an 
important role in quantum models \cite{newvertex, noi, loro}.}

To further study this correspondence, a 
canonical analysis of the action \Ref{action} is in progress, and will appear elsewhere \cite{Bianca2}.

\section{Conclusions}
We introduced a modified Regge calculus where the fundamental variables are areas and 3d dihedral angles
between triangles. The action, given in \Ref{action}, is the conventional Regge term 
with independent area-angle variables, plus two additional constraints. The first imposes the closure
of each tetrahedron in the triangulation, 
the second guarantees a consistent gluing between adjacent tetrahedra, by imposing
the (conformal) geometry of the common triangle to be the same.
All the constraints are local in the triangulation.
We expect our action to be related to a discretization of Plebanski's action.

Our main result is to show that the full set of constraints guarantees 
that local variables determine completely the geometry of each
4-simplex \emph{and of each tetrahedron} in the triangulation. As 4-simplices are glued together
identifying a tetrahedron in their boundary, being able to determine the tetrahedra's geometry is
crucial to have a consistent propagation of the degrees of freedom in the triangulation.
The crucial counting of the independent constraints was performed by linearizing the
constraints around generic non-degenerate configurations (see Appendix B). In particular we checked that
the potentially harmful orthogonal configuration is well behaved. 
Similarly, the ambiguity of two sets of lengths
giving the same areas \cite{Barrett} is removed in our formalism simply because the two sets give
different 3d angles. On the basis of our analysis, we can not exclude completely the presence
of pathological configurations. However, a well-behaved orthogonal configuration reassures us that
at least for the regular lattice our approach solves both difficulties with area Regge calculus described in the introduction. This opens the way, for instance, to perturbation theory on a flat background.

While studying the constraints we found a number of relations between the dihedral angles of various
dimension of a 4-simplex. We present them together with their derivation in the Appendix.
We also provide an explicit algorithm to compute the edge lengths from area-angles in
a tetrahedron and in a 4-simplex.

We expect our result to have a number of applications, and before concluding, 
we would like to briefly point out a few potentially promising ones.

At the classical level, the canonical analysis of \Ref{action} could shed light
on the description of the Hamiltonian algebra of deformation of discrete manifolds \cite{Bianca2}.
The applicability of this approach to numerical studies of lattice gravity has to be
explored. It would for instance be interesting to study whether our approach keeps the good
convergence properties of conventional Regge calculus in the continuum limit \cite{Brewin}.

At the quantum level, there are possible links to the spinfoam formalism that are worth exploring.
The formalism is expected to provide a well-defined measure for a regularized path integral
for non-perturbative quantum gravity (however see also \cite{Bojo}). 
Recently a spinfoam model has been proposed \cite{newvertex} (see also \cite{NewImmirzi}), 
whose dynamical variables can be expressed as normals to 
triangles \cite{noi} (see also \cite{loro}).
The scalar reduction of these quantities produces exactly the variables $(A_t, \phi_e^\tau)$
considered here. The matching of variables suggests that the discrete calculus introduced here is a candidate for
the semiclassical limit of this new spinfoam model, mimicking 
what happens in the 3d case with Regge calculus \cite{Ponzano}.
Indeed, the recent advances \cite{Ashtekar} on calculating the graviton propagator from spinfoams \cite{grav}
are based precisely on such a link. In the context of pure area Regge calculus on a single
4-simplex, this idea was investigated in \cite{Bianchi}.

From this viewpoint, it would be useful to study the quantum theory defined on a regular lattice
in perturbative expansion around flat spacetime, as done by Rocek and Williams \cite{Rocek}.

\subsubsection*{Acknowledgements}
We would like to thank Ruth M. Williams for her inspirational work
and Carlo Rovelli for useful discussions.
This research was supported by Perimeter Institute for Theoretical Physics.  
Research at Perimeter Institute is supported by the Government of Canada through 
Industry Canada and by the Province of Ontario through the Ministry of Research \& Innovation.

\appendix

\section{Relations between dihedral angles}
To study the geometry of simplices, we use the affine (or barycentric) coordinates  for Regge calculus 
introduced in \cite{Sorkin} and further developed in \cite{Reggae1} 
(see also \cite{bianca1}). With the help of these we derive relations between dihedral angles, 
and show how constrained areas and angles suffice to reconstruct edge lengths in 
$4$-simplices.\footnote{Given our interest in a Regge-like formulation of discrete geometry,
we focus here only on the case of flat simplices, with curvature concentrated on the $(n-2)$-subsimplices.
The same technique can be used to find similar relations in the case of spherical or hyperbolical $n$-simplices.}

Consider an $n$-dimensional simplex $\sigma$ with $n\geq 3$ and choose an $(n-3)$-d subsimplex $\sigma(ijk)$. (Here $\sigma(ijk\ldots)$ denotes the subsimplex spanned by all the vertices of $\sigma$ except for $(ijk\ldots)$.) We have three $(n-2)$-d subsimplices $\sigma(ij),\sigma(ik),\sigma(jk)$ meeting at $\sigma(ijk)$. These carry the three $n$-d dihedral angles $\th_{ij}$, $\th_{ik}$ and $\th_{jk}$ respectively. 

In addition, the subsimplex $\sigma(ijk)$ is shared by three $(n-1)$-d subsimplices $\sigma(i),\sigma(j)$ and $\sigma(k)$. In the intrinsic geometry of these subsimplices we can define the $(n-1)$-d angles $\phi_{lm,i}$, where $\phi_{lm,p}$ denotes the dihedral angle in the subsimplex $\sigma(p)$ between the simplices $\sigma(lp)$ and $\sigma(mp)$. There are again three $(n-1)$-d angles $\phi_{ij,k},\phi_{jk,i}$ and $\phi_{ik,j}$ meeting at the subsimplex $\sigma(ijk)$. 

It turns out that the set of angles $\th_{ij}$, $\th_{ik}$ and $\th_{jk}$ can be computed from the set of $(n-1)$-d angles $\phi_{ij,k},\phi_{jk,i}$ and $\phi_{ik,j}$ and vice versa. There are different methods to derive these relations. Here we use the affine metric, refering to the Appendix of \cite{Reggae1} for 
an introduction to affine coordinates and more details.

The affine metric $\tl g^{ij}$ associated to an $n$-simplex satisfies the following properties,
\equ\label{aff1}
\tl g^{ij} = -\f1{V^2}\,\f{\p V^2}{\p \ell_{ij}{}^2}, \quad i\neq j, \qquad
\tl g^{ii} = \f1{n^2}\,\f{V(i)^2}{V^2}.
\nequ
Given the normal $n(i)$ to a subsimplex $i$, we can use the affine metric to compute the scalar products
\eqa
n(i) \cdot n(i) &=& \tl g^{ii} \equiv |n(i)|^2, \\\label{nn}
n(i) \cdot n(j) &=& \tl g^{ij} \equiv - |n(i)| \, |n(j)| \, \cos\th_{ij}.
\neqa
Notice that with this conventions the closure condition reads
\eqa\nn
V(j)= \sum_{i\neq j} V(i)\cos\th_{ij}.
\neqa
In analogy with the continuum, the normal vectors allow us to introduce the induced
metric on a subsimplex $k$,
\equ\label{gind}
\tl g^{ij}(k) = \tl g^{ij} - \f{\tl g^{ik} \tl g^{jk}}{\tl g^{kk}}.
\nequ

From \Ref{nn} and (\ref{aff1}b) we obtain
\equ\label{theta1}
\cos\th_{ij} = - n^2 \f{V^2}{V(i) V(j)} \tl g^{ij}.
\nequ
This relation can be straighforwardly pushed one dimension down using the induced metric \Ref{gind}, to give
\eqa\label{ciccio}
\cos\phi_{ij,k} = - (n-1)^2 \f{V(k)^2}{V(ik) V(jk)} \tl g^{ij}(k). 
\neqa
Using \Ref{theta1} and the well-known generalized law of sines
\eqa\label{sines}
\sin\th_{ij}=\f{n}{(n-1)}\f{V(ij)V}{V(i)V(j)}
\neqa
to eliminate the volume factors in (\ref{ciccio}) we finally get
\eqa\label{phifromtheta}
\cos\phi_{ij,k}&=& \f{\cos\th_{ij}+ \cos\th_{ik} \cos\th_{jk}}{\sin\th_{ik} \sin\th_{jk}}. 
\neqa
These are relations between the three $\phi$ angles and the three $\theta$ angles at the subsimplex $\sigma(ikl)$.
The inverse, giving the $\theta$ angles as function of the three $\phi$ angles, has the remarkable
following form,
\eqa\label{thetafromphi}  
\cos\th_{ij}&=& \f{\cos\phi_{ij,k} - \cos\phi_{ik,j} \cos\phi_{jk,i}}{\sin\phi_{ik,j} \sin\phi_{jk,i}}. 
\neqa

The formulas (\ref{phifromtheta}) and (\ref{thetafromphi}) can be adapted to one dimension down, giving
the relations \Ref{a} and \Ref{ainv} in the main text between $(n-1)$- and $(n-2)$-dimensional 
angles.\footnote{An interesting alternative derivation of (\ref{phifromtheta}) and (\ref{thetafromphi}) 
uses a vanishing curvature condition. Consider the three $(n-1)$-dimensional subsimplices joined at $\sigma(ijk)$ as a piecewise linear $(n-1)$ dimensional geometry. Its intrinsic curvature 
is given by the deficit angle $\eps_{ij} = 2\pi - \phi_{ij,k} - \phi_{jk,i} - \phi_{ki,j}$, i.e. $2\pi$ minus the sum of the three $(n-1)$ dimensional angles meeting at $\sigma(ijk)$. In the flat $n$-dimensional embedding provided by
the $n$-simplex, its extrinsic curvature is given by the three $n$-dimensional dihedral angles.
That is if we parallel-transport a $n$-dimensional vector according to the $n$-dimensional geometry around $\sigma(ijk)$ we should obtain the identity. The latter is a rotation in the 3d subspace orthogonal to $\sigma(ijk)$
that can be expressed using the $\th$'s and $\phi$'s. Hence requiring this rotation to be equal to the identity leads to three conditions which can be used to express one set of angles as functions of the other set.} 

For the special case of 3d space, these formulas had already appeared in the mathematical (e.g. \cite{Luo}) and physical (e.g. \cite{'tHooft, Freidel, Reggae1}) literature.

From the formulas above, we can derive a number of other useful relations 
between areas and angles or between angles alone. Introducing the shorthand notation
$\rc(ij,k) \equiv \cos\phi_{ij,k}$, $\rs(ij,k) \equiv \sin\phi_{ij,k}$, we have
\eqa\nn
\frac{\rs(jl,i) \, \rs(ip,j)}{\rs(jp,i) \, \rs(il,j)} &=& \frac{V(ip) \, V(lj)}{V(il) \, V(pj)},
\\\nn\\\nn
\f{\rc(kl|i) - \rc(ki|l) \, \rc(li|k)}{\rs(ki|l) \, \rs(li|k)} &=& 
\f{\rc(kl|j) - \rc(kj|l) \, \rc(lj|k)}{\rs(kj|l) \, \rs(lj|k)},
\\\nn\\\nn
\rs(jl|i) \, \rs(kl|j) \, \rs(il|k) &=& \rs(kl|i) \, \rs(il|j) \, \rs(jl|k),
\neqa
as well as 
\eqa\label{quattro}
&& \rs(ij|k) \, \rs(ik|l) \, \rs(jl|i) \, \rs(kl|j) = \no && \hspace{0.5cm} 
= \rs(ij|l) \, \rs(ik|j) \, \rs(kl|i) \, \rs(jl|k).
\neqa
Any of these relations can be taken as the starting point to constrain the area-angle variables.
Their geometric interpretation is less immediate than the relations \Ref{aa}, and we do not discuss
them here. The interested reader can work them out easily.

\section{Lengths from area-angles}
In this Appendix, we show how to explicitly construct the edge lengths from area-angles in a tetrahedron
and in the full 4-simplex. Let us begin by considering the tetrahedron $i$, in the notation used so far.
The closure reads
\eqa
V(ik)= \sum_{j\neq i} V(jk)\cos\phi_{ij,k}.
\neqa
Heron's formula for the area gives 
\equ\nn
16 \, V(ij)^2 = 2 \sum_{k,l\neq i,j} V(ijk)^2 V(ijl)^2 - \sum_{k\neq i,j} V(ijk)^4.
\nequ
Using the law of sines \Ref{sines} adapted to 3d, 
the RHS gives $f_{ij}\big(V(ik), \sin\phi_{kl,i}\big)/{V^4}$ where $f_{ij}$
is a simple polynomial in the areas and the sines. 
From this we read an (asymmetric) expression for the tetrahedron's volume in terms of
areas and angles, 
\equ\label{vol}
V(i)^4 = \f1{16 V(ij)^2} \, f_{ij}\big(V(ik), \sin\phi_{kl,i}\big). 
\nequ 
Finally, we can use \Ref{sines} to express the six edge lengths
in terms of areas and angles, 
\equ\label{ltet}
V(ijk) = \f43 f_{ij}^{-\f14} {V(ij)^{\f32} V(ik)} \sin\phi_{jk,i}.
\nequ
Notice that the procedure becomes ill-defined for degenerate configurations where $V(i)=0$,
but it is perfectly well-defined otherwise, including orthogonal configurations with right angles.

For the 4-simplex we proceed exactly as above, using the relations \Ref{ltet} in each tetrahedron.
All that remains to prove is that the conditions \Ref{aa} are enough to ensure that \Ref{ltet}
applied to the different tetrahedra sharing the same edge gives the same value, namely that
\eqa\nn
V(ijk) &=& \f23 \f{V(ij) V(ik)}{V(i)} \sin\phi_{jk,i} = \\\nn && \hspace{-1.2cm}  = 
\f23 \f{V(ij) V(jk)}{V(j)} \sin\phi_{ik,j} = \f23 \f{V(ik) V(jk)}{V(k)} \sin\phi_{ij,k}.
\neqa
Thus for instance we need to show that
\equ\nn
\f{V(i)}{V(ik) \sin\phi_{jk,i}} = \f{V(j)}{V(jk) \sin\phi_{ik,j}}.
\nequ
This is a tedious but straighforward check that can be done using \Ref{vol}
for the two volumes, and then twice \Ref{quattro}.

This construction shows explicitly how the constraints \Ref{aa} and \Ref{closure},
which imply \Ref{quattro} and \Ref{vol}, allow to reconstruct unambiguously
the edge lengths of a tetrahedron and of a 4-simplex starting from areas and angles.

To make sure that the set of constraints \Ref{aa} and \Ref{closure} are sufficient, 
we studied the matrix of coefficients of their linearization in a 4-simplex.
The entries in this matrix depend on the configuration around which we choose to linearize.
We considered two different configurations, the equilateral and the orthogonal ones.
The equilateral one has no $\phi$ angles equal to $\pi/2$ (which would lead
to zeroes in the matrix of coefficients), and in this sense is a representative of
a generic configuration.
The orthogonal one has the maximal number of $\phi$ angles equal to $\pi/2$ 
and thus the maximal number of zeroes in the matrix of coefficients. It is the
dangerous configuration where area Regge calculus is ill-defined, and it
corresponds to a 4-simplex fitting into a regular hypercubical lattice.
In both cases the rank of the matrix turned out to be exactly 30, meaning that the initial
forty variables can be reduced to ten, which can then be chosen to be the edge lengths using
the explicit procedure described above.
The same procedure was used to check that \Ref{C1} has rank 21.

This shows that there cannot be singular configurations where two different sets of lengths
(and thus two different metrics) correspond to the same set of area-angle variables, as it is the
case for area Regge calculus.



\begin{thebibliography}{10}


\bibitem{Regge}
  T.~Regge,
  ``General relativity without coordinates,''
  Nuovo Cim.\  {\bf 19} (1961) 558.

\bibitem{RuthRegge}
T.~Regge and R.~M.~Williams,
  ``Discrete structures in gravity,''
  J.\ Math.\ Phys.\  {\bf 41}, 3964 (2000)
  [arXiv:gr-qc/0012035].
  
\bibitem{Gentle}
  R.~Loll,
  ``Discrete approaches to quantum gravity in four dimensions,''
  Living Rev.\ Rel.\  {\bf 1} (1998) 13
  [arXiv:gr-qc/9805049].
  A.~P.~Gentle and W.~A.~Miller,
  ``A brief review of Regge calculus in classical numerical relativity,''
  arXiv:gr-qc/0101028.
  H.~W.~Hamber,
  ``Discrete and Continuum Quantum Gravity,''
  arXiv:0704.2895 [hep-th].

\bibitem{Rocek}
  M.~Rocek and R.~M.~Williams,
  ``Quantum Regge Calculus,''
  Phys.\ Lett.\ B {\bf 104} (1981) 31.
 M.~Rocek and R.~M.~Williams,
  ``The Quantization Of Regge Calculus,''
  Z.\ Phys.\ C {\bf 21} (1984) 371.
  H.~W.~Hamber and R.~M.~Williams,
  ``Simplicial quantum gravity in three-dimensions: Analytical and numerical results,''
  Phys.\ Rev.\  D {\bf 47} (1993) 510.

\bibitem{Hamber}
  H.~W.~Hamber and R.~M.~Williams,
  ``Non-perturbative gravity and the spin of the lattice graviton,''
  Phys.\ Rev.\  D {\bf 70} (2004) 124007
  [arXiv:hep-th/0407039].

\bibitem{Immirzi}
  G.~Immirzi,
  ``Quantum gravity and Regge calculus,''
  Nucl.\ Phys.\ Proc.\ Suppl.\  {\bf 57} (1997) 65
  [arXiv:gr-qc/9701052].
     
\bibitem{carlo}
C.\,Rovelli,  \newblock {\em Quantum Gravity}  \newblock (Cambridge
University Press, Cambridge 2004). 

\bibitem{Plebanski}
  J.~F.~Plebanski,
  ``On the separation of Einsteinian substructures,''
  J.\ Math.\ Phys.\  {\bf 18}, 2511 (1977).
  
\bibitem{carloarea}
C.\,Rovelli: 
``The Basis of the Ponzano-Regge-Turaev-Viro-Ooguri quantum gravity model in
the loop representation basis'', 
Phys.\,Rev.\,{\bf D48} (1993) 2702. 


\bibitem{Makela}
  J.~Makela,
  ``On the phase space coordinates and the Hamiltonian constraint of Regge calculus,''
  Phys.\ Rev.\  D {\bf 49} (1994) 2882.
  J.~Makela,
  ``Variation of area variables in Regge calculus,''
  Class.\ Quant.\ Grav.\  {\bf 17}, 4991 (2000)
  [arXiv:gr-qc/9801022].
  J.~Makela and R.~M.~Williams,
  ``Constraints on area variables in Regge calculus,''
  Class.\ Quant.\ Grav.\  {\bf 18}, L43 (2001)
  [arXiv:gr-qc/0011006].

\bibitem{Barrett}
  J.~W.~Barrett, M.~Rocek and R.~M.~Williams,
  ``A note on area variables in Regge calculus,''
  Class.\ Quant.\ Grav.\  {\bf 16}, 1373 (1999)
  [arXiv:gr-qc/9710056].
  
\bibitem{Wainwright}
  C.~Wainwright and R.~M.~Williams,
  ``Area Regge calculus and discontinuous metrics,''
  Class.\ Quant.\ Grav.\  {\bf 21} (2004) 4865
  [arXiv:gr-qc/0405031].

\bibitem{Mike}
  M.~P.~Reisenberger,
  ``A left-handed simplicial action for euclidean general relativity,''
  Class.\ Quant.\ Grav.\  {\bf 14} (1997) 1753
  [arXiv:gr-qc/9609002].
  M.~P.~Reisenberger,
  ``Classical Euclidean general relativity from *left-handed area = right-handed area*,''
  arXiv:gr-qc/9804061.
  
\bibitem{CarloUn}  
C. Rovelli, \emph{unpublished} (2007). Some ideas then developed into the classical 
action used in \cite{newvertex}.

\bibitem{Barrett1}
  J.~W.~Barrett,
  ``First order Regge calculus,''
  Class.\ Quant.\ Grav.\  {\bf 11}, 2723 (1994)
  [arXiv:hep-th/9404124].

\bibitem{Freidel}
L.~Freidel and D.~Louapre,
``Asymptotics of 6j and 10j symbols,''
Class.\ Quant.\ Grav.\  {\bf 20} (2003) 1267
[arXiv:hep-th/0209134].

\bibitem{Caselle}
  M.~Caselle, A.~D'Adda and L.~Magnea,
  ``Regge calculus as a local theory of the Poincare group,''
  Phys.\ Lett.\  B {\bf 232} (1989) 457.

\bibitem{Pleb}
  R.~Capovilla, T.~Jacobson and J.~Dell,
  ``General Relativity Without the Metric,''
  Phys.\ Rev.\ Lett.\  {\bf 63}, 2325 (1989).
  R.~De Pietri and L.~Freidel,
  ``so(4) Plebanski Action and Relativistic Spin Foam Model,''
  Class.\ Quant.\ Grav.\  {\bf 16} (1999) 2187
  [arXiv:gr-qc/9804071].
    A.~Perez,
  ``Spin foam quantization of SO(4) Plebanski's action,''
  Adv.\ Theor.\ Math.\ Phys.\  {\bf 5}, 947 (2002)
  [Erratum-ibid.\  {\bf 6}, 593 (2003)]
  [arXiv:gr-qc/0203058].
  L.~Smolin,
  ``The Plebanski action extended to a unification of gravity and Yang-Mills theory,''
  arXiv:0712.0977 [hep-th].


\bibitem{newvertex}
  J.~Engle, R.~Pereira and C.~Rovelli,
  ``The loop-quantum-gravity vertex-amplitude,''
  Phys.\ Rev.\ Lett.\  {\bf 99}, 161301 (2007)
  [arXiv:0705.2388 [gr-qc]].
  J.~Engle, R.~Pereira and C.~Rovelli,
  ``Flipped spinfoam vertex and loop gravity,''
  Nucl.\ Phys.\  B {\bf 798}, 251 (2008)
  [arXiv:0708.1236 [gr-qc]].

\bibitem{noi}
 E.~R.~Livine and S.~Speziale,
``New spinfoam vertex for quantum gravity",
  Phys.\ Rev.\  D {\bf 76} (2007) 084028
  [arXiv:0705.0674 [gr-qc]].
  E.~R.~Livine and S.~Speziale,
  ``Solving the Simplicity Constraints for Spinfoam Quantum Gravity,''
  Europhys.\ Lett.\  {\bf 81} (2008) 50004
  [arXiv:0708.1915 [gr-qc]].

\bibitem{loro}
  L.~Freidel and K.~Krasnov,
  ``A New Spin Foam Model for 4d Gravity,''
  arXiv:0708.1595 [gr-qc].


\bibitem{Bianca2}
B. Dittrich and J. P. Ryan, ``Phase space descriptions for 4d simplicial geometries,''
arXiv:0807.2806 [gr-qc].

\bibitem{Brewin}
  L.~C.~Brewin and A.~P.~Gentle,
  ``On the convergence of Regge calculus to general relativity,''
  Class.\ Quant.\ Grav.\  {\bf 18} (2001) 517
  [arXiv:gr-qc/0006017].

\bibitem{Bojo}
  M.~Bojowald and A.~Perez,
  ``Spin foam quantization and anomalies,''
  arXiv:gr-qc/0303026.

\bibitem{NewImmirzi}
J.~Engle, E.~Livine, R.~Pereira and C.~Rovelli,
  ``LQG vertex with finite Immirzi parameter,''
  arXiv:0711.0146 [gr-qc].
  
\bibitem{Ponzano}
G.~Ponzano, T.~Regge.  ``Semiclassical limit of
Racah coefficients", in {\em Spectroscopy and group theoretical methods
in Physics}, F.~Bloch ed. (North-Holland, Amsterdam, 1968).

\bibitem{Ashtekar}
  A.~Ashtekar,
  ``Loop quantum gravity: Four recent advances and a dozen frequently asked questions,''
  arXiv:0705.2222 [gr-qc].

\bibitem{grav}
   C.~Rovelli,
  ``Graviton propagator from background-independent quantum gravity,''
  Phys.\ Rev.\ Lett.\  {\bf 97} (2006) 151301
  [arXiv:gr-qc/0508124].
  S.~Speziale,
  ``Towards the graviton from spinfoams: The 3d toy model,''
  JHEP {\bf 05} (2006) 039 [arXiv:gr-qc/0512102].
 E.~Bianchi, L.~Modesto, C.~Rovelli and S.~Speziale,
  ``Graviton propagator in loop quantum gravity,''
    Class.\ Quant.\ Grav.\  {\bf 23} (2006) 6989
  [arXiv:gr-qc/0604044].
  E.~R.~Livine, S.~Speziale and J.~L.~Willis,
  ``Towards the graviton from spinfoams: Higher order corrections in the 3d
  toy model,''
  Phys.\ Rev.\  D {\bf 75} (2007) 024038
  [arXiv:gr-qc/0605123].
  E.~R.~Livine and S.~Speziale,
  ``Group integral techniques for the spinfoam graviton propagator,''
  JHEP {\bf 0611} (2006) 092
  [arXiv:gr-qc/0608131].
  J.~D.~Christensen, E.~R.~Livine and S.~Speziale,
  ``Numerical evidence of regularized correlations in spin foam gravity,''
  arXiv:0710.0617 [gr-qc].
  E.~Alesci and C.~Rovelli,
  ``The complete LQG propagator: I. Difficulties with the Barrett-Crane vertex,''
  Phys.\ Rev.\  D {\bf 76} (2007) 104012
  [arXiv:0708.0883 [gr-qc]].
  E.~Alesci and C.~Rovelli,
  ``The complete LQG propagator: II. Asymptotic behavior of the vertex,''
  Phys.\ Rev.\  D {\bf 77} (2008) 044024
  [arXiv:0711.1284 [gr-qc]].


\bibitem{Bianchi}
  E.~Bianchi and L.~Modesto,
  ``The perturbative Regge-calculus regime of Loop Quantum Gravity,''
  arXiv:0709.2051 [gr-qc].

\bibitem{Sorkin}
  R.~Sorkin,
  ``The Electromagnetic Field On A Simplicial Net,''
  J.\ Math.\ Phys.\  {\bf 16} (1975) 2432
  [Erratum-ibid.\  {\bf 19} (1978) 1800].
    
\bibitem{Reggae1}
  B.~Dittrich, L.~Freidel and S.~Speziale,
  ``Linearized dynamics from the 4-simplex Regge action,''
  Phys.\ Rev.\  D {\bf 76}, 104020 (2007)
  [arXiv:0707.4513 [gr-qc]].

\bibitem{bianca1}
  B.~Dittrich and R.~Loll,
  ``Counting a black hole in Lorentzian product triangulations,''
  Class.\ Quant.\ Grav.\  {\bf 23} (2006) 3849
  [arXiv:gr-qc/0506035].
    
\bibitem{Luo}
Feng Luo, ``Volume and angle structures on 3-manifolds,'' [arXiv:math/0504049].

\bibitem{'tHooft}
  G.~'t Hooft,
  ``Canonical quantization of gravitating point particles in (2+1)-dimensions,''
  Class.\ Quant.\ Grav.\  {\bf 10}, 1653 (1993)
  [arXiv:gr-qc/9305008].
  
  

\end{thebibliography}
\end{document}